\documentclass[conference]{IEEEtran}

\usepackage{cite}
\usepackage{amsmath,amssymb,amsfonts}
\usepackage{algorithm,algpseudocode,algorithmicx}
\usepackage{graphicx}
\usepackage{textcomp}
\usepackage{xcolor}
\usepackage{tikz}

\usepackage{bm} 
\usepackage{booktabs} 
\usepackage{multirow} 
\usepackage{stmaryrd} 

\usepackage{soul}  
\sethlcolor{yellow} 
\definecolor{lightyellow}{RGB}{255, 255, 197} 
\definecolor{lightred}{RGB}{255, 220, 209} 

\usepackage{arydshln} 
\begin{document}

\title{FAME: \underline{F}PGA \underline{A}cceleration of Secure \underline{M}atrix Multiplication with Homomorphic \underline{E}ncryption}

\author{\IEEEauthorblockN{Zhihan Xu$^{*}$, Rajgopal Kannan$^{\dagger}$ and Viktor K. Prasanna$^{*}$}
 \IEEEauthorblockA{$^{*}$University of Southern California, USA\\
$^{\dagger}$DEVCOM Army Research Office 
\\
Email: \{\mbox{zhihanxu, prasanna\}@usc.edu}, rajgopal.kannan.civ@army.mil}}

\maketitle


\begin{abstract}
Homomorphic Encryption (HE) enables secure computation on encrypted data, addressing privacy concerns in cloud computing. However, the high computational cost of HE operations, particularly matrix multiplication (MM), remains a major barrier to its practical deployment. Accelerating Homomorphic Encrypted MM (HE MM) is crucial for applications such as privacy-preserving machine learning.

In this paper, we present a bandwidth-efficient FPGA implementation of HE MM. We first develop a cost model to evaluate the on-chip memory requirement for a given set of HE parameters and input matrix sizes. 
Our analysis shows that optimizing on-chip memory usage is critical for scalable and efficient HE MM.
To this end, we design a novel datapath for Homomorphic Linear Transformation (HLT), the major bottleneck in HE MM. Our datapath significantly reduces off-chip memory traffic and on-chip memory demand by enabling fine-grained data reuse.
Leveraging the proposed datapath, we introduce FAME, the first FPGA-based accelerator specifically tailored for HE MM. 
FAME supports arbitrary matrix shapes and is configurable for a wide range of HE parameter sets.
We implement FAME on Alveo U280 and evaluate its performance over diverse matrix sizes and shapes. Experimental results show that FAME achieves an average of 221$\times$ speedup over state-of-the-art CPU-based implementations, demonstrating its scalability and practicality for large-scale consecutive HE MM and real-world workloads.

\end{abstract}

\begin{IEEEkeywords}
FPGA Accelerator, Secure Matrix Multiplication, Homomorphic Encryption
\end{IEEEkeywords}

\section{Introduction}~\label{sec:intro}
Cloud computing has revolutionized many applications across various sectors such as healthcare, finance, and government~\cite{varghese2018next,vasiljeva2017cloud}. Despite its computational advantages, security and data privacy remain concerns in adopting cloud computing services~\cite{cloud_sec}. Homomorphic Encryption (HE) offers a promising solution by enabling computations directly on encrypted data without decrypting the data at the server side. However, HE introduces substantial computational overhead, often several orders of magnitude higher than equivalent plaintext operations~\cite{de2021does}. As a result, accelerating HE applications has become a focal point of recent research efforts~\cite{reagen2021cheetah,ebel2023orion,zhu2023fxhenn,samardzic2022craterlake,ye2021performance,yang2022fpga,ren2023cham,yang2022bandwidth}.


Existing HE acceleration efforts~\cite{reagen2021cheetah,ebel2023orion,zhu2023fxhenn,samardzic2022craterlake,ye2021performance,yang2022fpga,ren2023cham,yang2022bandwidth} primarily focus on scenarios where the machine learning (ML) model remains in plaintext and only the input data is encrypted. For example, works such as~\cite{yang2022fpga,ren2023cham,yang2022bandwidth} accelerate matrix-vector multiplication where the matrix (i.e., the model) is unencrypted. While this setup reduces computational cost, it is not sufficient for applications requiring stronger privacy guarantees, where both the model and data must remain encrypted.
Two representative scenarios illustrate this need: (1) a cloud service provider owns the model, trains the model with the encrypted data from the data owner, and uses the encrypted trained model to make inferences on new encrypted inputs; (2) a third-party model provider (e.g., researchers or companies) owns the trained encrypted model and uploads it to the cloud service provider to perform inference on encrypted inputs from various data owners. In both scenarios, both the model and the input data are encrypted, necessitating secure and efficient HE matrix multiplication (MM) to support fully encrypted inference workflows.

HE MM is defined as matrix multiplication in which both input matrices are encrypted with homomorphic encryption. The result of HE MM is encrypted output matrix.
HE MM is a fundamental kernel in many HE applications. Several algorithmic advancements~\cite{e2dm,huang2021more,huang2023secure,gao2024secure} to reduce its computational complexity have been made. However, CPU-based implementations remain prohibitively slow, particularly for large matrix sizes. For example, multiplying two modestly sized matrices (e.g., $64 \times 64$) under a small HE parameter set can still take several tens of seconds~\cite{huang2023secure}, which is approximately 1000$\times$ slower than the plaintext operation. This substantial performance gap highlights the need for hardware acceleration to make HE MM based applications practical.


FPGAs offer a unique advantage for accelerating HE MM due to their configurable on-chip memory hierarchy and customizable compute resources, enabling fine-grained datapath control to enhance on-chip data reuse. In addition, leading FPGA vendors also incorporate High Bandwidth Memory (HBM) to overcome traditional DRAM bandwidth limitation, making them well-suited for handling the high-throughput demands of encrypted data transfers. 

The primary bottleneck in HE MM is homomorphic linear transformation (HLT), accounting for over 95\% of the total runtime~\cite{e2dm}. This inefficiency arises from two key challenges: (1) HLT requires multiple ciphertext rotations - one of the most computationally expensive operations in HE, and (2) it generates numerous intermediate ciphertexts during computation, leading to substantial off-chip memory traffic that limits the scalability of HE MM to larger matrices and HE parameter sets.
To mitigate the first challenge, a pioneering work~\cite{halevi2018faster} introduced 
the \textit{hoisting} technique to amortize certain sub-operations across multiple rotations, which has been widely adopted in subsequent works~\cite{cheon2019faster,han2019improved,de2021does,xu2024bandwidth}.
However, the second challenge remains unaddressed due to the lack of hardware-aware solutions. In this work, we address both challenges by designing a novel HLT datapath that integrates algorithmic optimizations to minimize off-chip memory traffic. Our key contributions are summarized as follows:
\begin{itemize}
    \item We develop a cost model to analyze the on-chip memory requirements of HE MM for a given matrix size, shape, and HE parameter set. The analysis shows that HE MM is not scalable without explicit datapath optimizations due to substantial off-chip ciphertext traffic.
    \item We propose an on-chip memory-optimized datapath for HLT, the bottleneck in HE MM, through hardware-software co-design. The datapath significantly reduces off-chip memory traffic by fusing sub-operations and enabling fine-grained on-chip ciphertext reuse.
    \item Leveraging this datapath, we design FAME, the first accelerator tailored for HE MM. FAME achieves resource efficiency through reduced on-chip memory usage. The architecture can be configured to support diverse matrix sizes, shapes, and HE parameter sets.
    \item We implement FAME on Alveo U280. Experimental results show that FAME achieves $221\times$ speedup on average over state-of-the-art CPU implementations across diverse matrix sizes, shapes, and HE parameters, demonstrating its scalability and practicality for large-scale HE MM.
\end{itemize}




\section{Background}

\subsection{Threat Model}
We assume a semi-honest (honest but curious) cloud service provider as previous works~\cite{ran2022cryptogcn,juvekar2018gazelle,moon2024thor,yang2022bandwidth}. The client sends the encrypted input matrix to the cloud server to perform HE MM. Importantly, the other input matrix is also encrypted, whether it comes from the client, a third-party model provider, or the server itself. This threat model differs from prior HE acceleration works~\cite{yang2022bandwidth,ren2023cham,yang2022fpga}, which typically assume only two parties (client and server) and a plaintext model. In contrast, our approach addresses scenarios with stronger privacy guarantees by ensuring both inputs remain encrypted throughout the computation.

\subsection{Homomorphic Encryption (HE)}\label{sec:bg-he}
Among various HE schemes~\cite{brakerski2012fully,brakerski2014leveled,fan2012somewhat,cheon2017homomorphic}, we focus on the CKKS~\cite{cheon2017homomorphic}, which supports real-number computations, making it suitable for HE machine learning (ML) applications.
In CKKS, a message vector $\bm{m}$ of $N/2$ numbers is first encoded into a plaintext (Pt) polynomial of degree $N-1$. The Pt is then encrypted into a ciphertext (Ct) $\llbracket\bm{m}\rrbracket:=(\bm{a},\bm{b})$, where both $\bm{a}$ and $\bm{b}$ are polynomials of degree $N-1$.
Each Ct polynomial resides in the ring $\mathcal{R}_Q = \mathbb{Z}_Q[X]/(X^N + 1)$, meaning the polynomial coefficients are integers modulo $Q$.

\subsubsection{Residual Number System (RNS)}
The full modulus $Q$ is typically hundreds to thousands of bits. To enable practical arithmetic, the RNS is used to decompose  $Q$ or $Q_L$ into the product of the coprime moduli $Q_L=\prod_{i=0}^{L}q_i$, with each $q_i$ fitting within a machine word. As a result, each ciphertext polynomial is split into $L+1$ limbs, where each limb is a sub-polynomial modulo $q_i$: $[\bm{a}]_{Q_L} \mapsto [\bm{a}]_{q_0},\cdots, [\bm{a}]_{q_L}$. A freshly encrypted Ct (i.e., a fresh Ct) consists of $L+1$ such limbs.


\subsubsection{CKKS Operations}
Assume a Pt encodes a message vector $\bm{m_p}$. CKKS supports the following operations based on modular arithmetic over encrypted vectors.
\begin{itemize}
    \item $\mathsf{Add}(\llbracket\bm{m}\rrbracket,\llbracket\bm{m}'\rrbracket)\rightarrow\llbracket\bm{m}+\bm{m}'\rrbracket=(\bm{a}+\bm{a}',\bm{b}+\bm{b}')$
     \item $\mathsf{Mult}(\llbracket\bm{m}\rrbracket,\llbracket\bm{m}'\rrbracket)\rightarrow\llbracket\bm{m}\odot\bm{m}'\rrbracket=(\bm{a}\odot \bm{a}',\bm{a}\odot \bm{b}'+\bm{b}\odot \bm{a}')+\mathsf{KeySwitch}(\bm{b}\odot \bm{b}', \textbf{evk}_{\text{mult}})$
     \item $\mathsf{CMult}(\llbracket\bm{m}\rrbracket, \text{Pt})\rightarrow\llbracket\bm{m}\odot\bm{m_p}\rrbracket=(\bm{a}\odot\text{Pt},\bm{b}\odot\text{Pt})$
     \item $\mathsf{Rot}(\llbracket\bm{m}\rrbracket,r)\rightarrow\llbracket\rho(\bm{m};r)\rrbracket=\mathsf{KeySwitch}(\psi_r(\bm{b}),\textbf{evk}_{\text{rot}}^{(r)}) \\ +(\psi_r(\bm{a}),\textbf{0})$, where $\rho(\bm{m};r)$ denotes a circular left rotation by $r$ slots, and $\psi_r$ is the automorphism ($\mathsf{Automorph}$) on a polynomial (i.e., a coefficient permutation). For the index mapping of $\psi_r$, please refer to~\cite {kim2022ark}.
\end{itemize}

\subsubsection{Sub-operations}
$\mathsf{Mult}$ and $\mathsf{Rot}$ rely on the computationally intensive $\mathsf{KeySwitch}$ sub-operation, involving an inner product ($\mathsf{KeyIP}$) between a Ct polynomial and an evaluation key $\mathbf{evk}$. The $\mathbf{evk}$ is a $2\times\beta$ matrix of polynomials modulo $PQ$, where $P=\prod_{i=0}^{k-1}p_i$ is the auxiliary modulus. 
To perform $\mathsf{KeyIP}$, the Ct polynomial is first decomposed ($\mathsf{Decomp}$) into $\beta$ digits, each comprising $\alpha:=(L+1)/\beta$ limbs. Each digit is then raised in modulus ($\mathsf{ModUp}$) from $Q_{\alpha+1}$ ($\alpha+1$ limbs) to $PQ_L$ ($L+k+1$ limbs). The $\mathsf{KeyIP}$ is computed by multiplying the digit vector with each row of $\mathbf{evk}$ matrix and accumulating along the row. This results in two polynomials modulo $PQ_L$. Finally, the result is reverted to modulus $Q_L$ using $\mathsf{ModDown}$. For details of $\mathsf{KeySwitch}$, we refer~\cite{cheon2019full,de2021does,han2020better}.


Number Theoretic Transform ($\mathsf{NTT}$), a finite-field variant of the Discrete Fourier Transform~\cite{ntt_survey}, enables element-wise polynomial multiplication in the evaluation domain. To avoid unnecessary $\mathsf{NTT}$s and its inversion $\mathsf{iNTT}$s, polynomials typically remain in the evaluation domain~\cite{kim2022ark,100x,kim2022bts}. However, $\mathsf{ModUp}$ and $\mathsf{ModDown}$ require base conversion ($\mathsf{BaseConv}$) to change the modulus (number of limbs), which cannot be performed in the evaluation domain. Therefore, $\mathsf{iNTT}$ is applied beforehand, and $\mathsf{NTT}$ is applied afterward.

\subsubsection{Ct Levels and Rescaling}
Each $\mathsf{Mult}$ or $\mathsf{CMult}$ is followed by a $\mathsf{Rescale}$, which is a special case of $\mathsf{ModDown}$, dropping the last limb ($q_L$) of the Ct polynomial~\cite{100x,kim2022ark}.
After $L$ such operations, only the base limb ($q_0$) remains, and further multiplications are not allowed. Thus, $L$ defines the maximum (or initial) Ct level, which supports an evaluation circuit with a computation depth of $L$. We use $\ell$ to denote the current Ct level, replacing $L$ when the Ct is not fresh.

\subsection{Secure Matrix Multiplication with HE}
In this work, we consider a general method for HE MM with arbitrary matrix shapes.
Given $A_{m\times l}\times B_{l\times n}$, state-of-the-art (SOTA) approaches~\cite{e2dm,huang2021more,gao2024secure,huang2023secure} rely on element-wise multiplication, where each matrix is encrypted into a Ct. We follow the notations from~\cite{gao2024secure}, as shown in Eq.~\ref{eq:mm}.
\begin{equation}
    A_{m \times l} \times B_{l \times n} = 
\sum_{k=0}^{l-1} \left( \epsilon^{k} \circ \sigma(A) \right) \odot 
\left( \omega^{k} \circ \tau(B) \right)
\label{eq:mm}
\end{equation}
where $\circ$ denotes composition. The four matrix transformation operators are defined as:
\begin{equation}
    \sigma(A)_{i,j} = A_{i, [i+j]_l}, \quad 0 \leq i < m, 0 \leq j < l
\end{equation}
\begin{equation}
\tau(B)_{i,j} = B_{[i+j]_l, j}, \quad 0 \leq i < l, 0 \leq j < n
\end{equation}
\begin{equation}
    \epsilon^{k}(A)_{i,j} = A_{i, [j+k]_l}, \quad 0 \leq i < m, 0 \leq j < n
\end{equation}
\begin{equation}
\omega^{k}(B)_{i,j} = B_{[i+k]_l, j}, \quad 0 \leq i < m, 0 \leq j < n 
\end{equation}

\begin{figure}
    \centering
    \includegraphics[width=\linewidth]{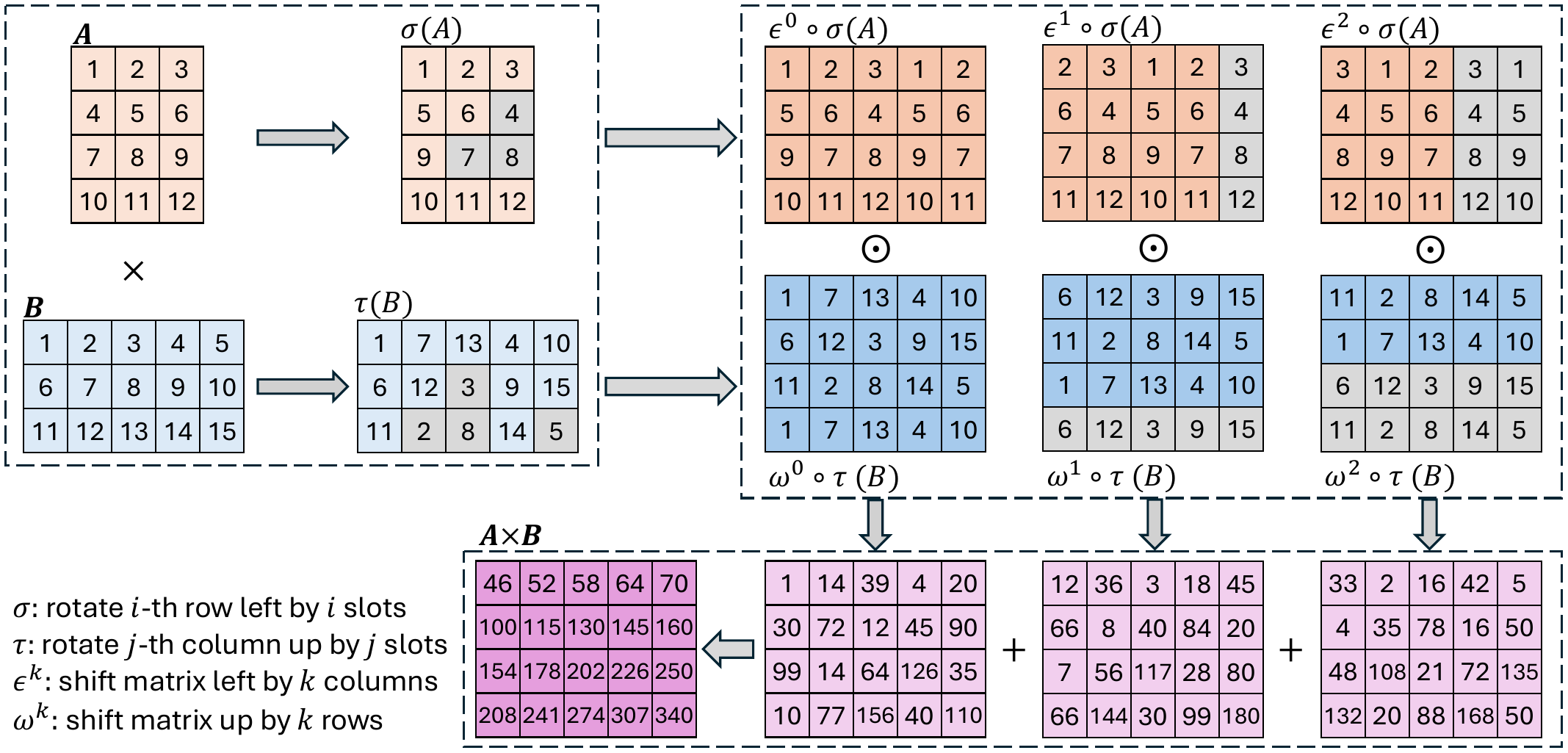}
    \caption{Element-wise MM with an example of $m=4,l=3,n=5$}
    \label{fig:mm}
\end{figure}

Fig.~\ref{fig:mm} illustrates an example of $A_{4\times 3}\times B_{3\times 5}$. With HE, each input matrix is flattened into a vector $\bm{m}$. 
Now, the four matrix transformations can be expressed as linear transformations $\bm{m}\rightarrow U\cdot \bm{m}$, where $U$ is the corresponding transformation matrix.
Assuming both $A$ and $B$ are flattened in the column-major order, the four transformation matrices are:
\begin{equation}
\begin{aligned}
    U^{\sigma}_{i + j \cdot m, h} &= 
    \begin{cases}
        1 & \text{if } h = i + [i + j]_l \cdot m \\
        0 & \text{otherwise}
    \end{cases} 
\end{aligned}
\end{equation}
\begin{equation}
\begin{aligned}
    U^{\tau}_{i + j \cdot l, h} &= 
    \begin{cases}
        1 & \text{if } h = [i + j]_l + j \cdot l \\
        0 & \text{otherwise}
    \end{cases} 
\end{aligned}
\end{equation}
\begin{equation}
\begin{aligned}
    U^{\epsilon^k}_{i, j} &= 
    \begin{cases}
        1 & \text{if } j = [k \cdot m + i]_{m \cdot l} \\
        0 & \text{otherwise}
    \end{cases} \\ 
\end{aligned}
\end{equation}
\begin{equation}
\begin{aligned}
    U^{\omega^k}_{i, j} &= 
    \begin{cases}
        1 & \text{if } j = [k + [i]_m]_l + \lfloor i / m \rfloor \cdot l \\
        0 & \text{otherwise}
    \end{cases}
\end{aligned}
\end{equation}

The four transformation matrices are sparse, each with multiple non-zero diagonals. For $U\in \mathbb{Z}_{x\times y}$ and $-x<z<y$, the $z$-th diagonal is defined as:
\begin{equation}
\begin{aligned}
    \bm{u}_z &= 
    \begin{cases}
        (U_{0,z}, U_{1,z+1},\cdots,U_{y-1-z,y-1},0,\cdots,0) & \text{if } z>0 \\
        (0,\cdots,0,U_{-z,0},U_{1-z,1},\cdots,U_{x-1,x-1-z}) & \text{if } z<0
    \end{cases}
\end{aligned}
\end{equation}
If $U$ has $d$ non-zero diagonals indexed by $z_0,\cdots, z_{d-1}$, the matrix-vector product can be computed using a combination of rotation and element-wise multiplication:
\begin{equation}
    U\cdot \bm{m}=\sum_{0\leq t<d}(\bm{u}_{z_t} \odot \rho(\bm{m}; z_t))
\end{equation}

Since both $A$ and $B$ are encrypted, the linear transformations must be performed homomorphically, referred to as $\mathsf{HLT}$ shown in Algorithm~\ref{algo:hlt}. Each $\mathsf{HLT}$ includes $\mathsf{CMult}$, which requires a $\mathsf{Rescale}$ at the end, reducing the Ct level by one. The MM in Eq.~\ref{eq:mm} is described in Algorithm~\ref{algo:hemm} with HE. 

\begin{algorithm}[t]
\caption{Homomorphic Linear Transformation}
\label{algo:hlt}
\textbf{Input:} Encrypted matrix: Ct, Pre-computed non-zero Pt diagonals of $U$ and their indices: \{$\bm{u}_{z_t},z_t$\}  \\
\textbf{Output:} $\text{Ct}'$ (initialized as zero) $\leftarrow\mathsf{HLT}(\text{Ct}, U)$
\begin{algorithmic}[1]
    \For{$t = 0$ to $d - 1$}
        \State $\text{Ct}' \gets \mathsf{Add}(\text{Ct}', \mathsf{CMult}(\mathsf{Rot}(\text{Ct}; z_{t}); \bm{u}_{z_{t}}))$
    \EndFor
    \State \textbf{return} $\mathsf{Rescale}(\text{Ct}')$
\end{algorithmic}
\end{algorithm}

\begin{algorithm}[t]
\caption{Homomorphic Encrypted Matrix Multiplication}
\label{algo:hemm} 
\textbf{Input:} $\text{Ct}_A,\text{Ct}_B,U^\sigma,U^\tau,U^{\epsilon^k},U^{\omega^k}$ 
\begin{algorithmic}[1]
    \State $\text{Ct}_{A^{(0)}}\leftarrow\mathsf{HLT}(\text{Ct}_A, U^\sigma)$
    \State $\text{Ct}_{B^{(0)}}\leftarrow\mathsf{HLT}(\text{Ct}_B, U^\tau)$
    \For{$k = 0$ to $l-1$}
        \State $\text{Ct}_{A^{(k+1)}}\leftarrow\mathsf{HLT}(\text{Ct}_{A^{(0)}},U^{\epsilon^k})$
        \State $\text{Ct}_{B^{(k+1)}}\leftarrow\mathsf{HLT}(\text{Ct}_{B^{(0)}},U^{\omega^k})$ 
        \State $\text{Ct}_{AB}\leftarrow\mathsf{Add}(\text{Ct}_{AB},\mathsf{Mult}(\text{Ct}_{A^{(k+1)}},\text{Ct}_{B^{(k+1)}}))$
    \EndFor
    \State \textbf{return} $\text{Ct}_{AB}$
\end{algorithmic}
\end{algorithm}





\section{Complexity Analysis and Cost Model}
In this section, we analyze the computational complexity of HE MM and introduce a cost model for estimating the on-chip memory required for Ct in performing HE MM.
\subsection{Complexity Analysis}\label{sec:algo}
Algorithm~\ref{algo:hemm} has two steps:
Step~1 (Lines 1–2): Generate $\text{Ct}_{A^{(0)}}$ and $\text{Ct}_{B^{(0)}}$.
Step~2 (Lines 3–6): For each $k = 0$ to $l-1$, generate $\text{Ct}_{A^{(k+1)}}$ and $\text{Ct}_{B^{(k+1)}}$, and perform a multiply-accumulate operation to produce the final encrypted result.

In total, HE MM requires $2 \cdot (l + 1)$ $\mathsf{HLT}$ operations. However, the cost of each $\mathsf{HLT}$ varies with the number of non-zero diagonals ($d$) in the associated transformation matrix, as this directly impacts the number of required $\mathsf{Rot}$. The values of $d$ for each type of transformation matrix are as follows~\cite{gao2024secure}:
\begin{equation}
    d_{U^\sigma}=2\cdot\text{min}(m,l)-1
\end{equation}
\begin{equation}
    d_{U^\tau}=2\cdot\text{min}(n,l)-1
\end{equation}
\begin{equation}
    d_{U^{\epsilon^k}}=\lfloor\frac{n}{l}\rfloor+1
\end{equation}
\begin{equation}
\begin{aligned}
    d_{U^{\omega^k}}=
    \begin{cases}
        2& \quad \text{if}\quad m=l \\
        n\cdot(\lfloor\frac{m}{l}\rfloor+2) & \quad \text{otherwise}
    \end{cases}
\end{aligned}
\end{equation}

The complexity of general HE MM is summarized in Table~\ref{tab:complexity}. Each $\mathsf{HLT}$ reduces the Ct level by one due to the required $\mathsf{Rescale}$. Additionally, the $\mathsf{Mult}$ operation in Step~2 incurs one more level reduction. Therefore, evaluating a single HE MM requires the Ct to have at least four levels, i.e., $L\geq4$.

\begin{table}[t]
    \centering
    \caption{Complexity and Required Depth of General HE MM}
    \label{tab:complexity}
    \begin{tabular}{c|c@{\hskip 10pt}c@{\hskip 10pt}c@{\hskip 10pt}c@{\hskip 10pt}c}
        \toprule
                &  $\mathsf{Add}$ & $\mathsf{Mult}$ & $\mathsf{CMult}$ & $\mathsf{Rot}$ & Depth$^*$\\
        \hline
        Step 1  &  $\phi$ & 0 & $\phi$ & $\phi$ & 1\\ 
        
        Step 2 & $\zeta+l$ & $l$ & $\zeta$ & $\zeta$ & 2 \\[1ex]
        
        Total& $\phi+\zeta+l$ & $l$ & $\phi+\zeta$ & $\phi+\zeta$  & 3\\
        \bottomrule
        \multicolumn{6}{l}{$\phi=d_{U^\sigma}+d_{U^\tau}; \zeta=l\cdot(d_{U^{\epsilon^k}}+d_{U^{\omega^k}})$}\\
        \multicolumn{6}{l}{$*$ Consumed Ct levels}
    \end{tabular}
\end{table}

\subsection{Cost Model}
Given the size of input matrices and HE parameters, the model estimates the on-chip memory needed to store all intermediate Cts on the chip while performing HE MM.
Since each Ct encrypts a vector of size $N/2$, the minimal polynomial degree $N$ must be large enough to encode each matrix ($A_{m\times l}$ and $B_{l\times n}$). Thus, $N$ is given by:
\begin{equation}\label{eq:degree}
    N=\text{max}(2^{\lceil\log_2(2ml)\rceil},2^{\lceil\log_2(2nl)\rceil}),
\end{equation}


\subsubsection{Data Sizes}
Let $\mathcal{B}_{\text{coeff}}$ denote the size (in bytes) of a single limb coefficient. Then, the size of one limb is given by $\mathcal{B}_{\text{limb}} = N \cdot \mathcal{B}_{\text{coeff}}$. The size of a Ct at level $\ell$ is:
\begin{equation}
\mathcal{B}_{\text{Ct}}^{\ell}=2\cdot(\ell+1)\cdot\mathcal{B}_{\text{limb}}=2\cdot N\cdot\log Q_\ell/8
\end{equation}

Each encoded Pt has the size of $\mathcal{B}_{\text{limb}}$. The $\mathsf{HLT}$ requires Pt diagonals of $U$, so the total Pt size is proportional to the number of non-zero diagonals.
Additionally, each $\mathsf{Rot}$ and $\mathsf{Mult}$ operation requires an $\mathbf{evk}$, with its size defined in Eq.~\ref{eq:key}. 
Moreover, the $\mathsf{NTT}$ and $\mathsf{iNTT}$ require additional twiddle factors, which are proportional to limb count. 
\begin{equation}\label{eq:key} \mathcal{B}_{\mathbf{evk}}^{\ell} = 2 \cdot \beta \cdot (\ell + k + 1) \cdot \mathcal{B}_{\text{limb}} \end{equation}



\subsubsection{On-chip Memory Requirement}\label{sec:sram}
As Pt diagonals, $\mathbf{evk}$, and twiddle factors are read-only after being fetched, we consider the on-chip memory requirement for Ct to explore on-chip Ct reuse.
In addition to storing two Cts of input matrices, we need to consider the intermediate Cts generated during HE MM. 
For simplicity, we assume all Cts are fresh (at maximum level $L$) and ignore Ct level reductions in our analysis, providing an upper bound on memory requirement.

The primary bottleneck is the $\mathsf{KeySwitch}$, required by both $\mathsf{Rot}$ and $\mathsf{Mult}$, with $\mathsf{Rot}$ being heavily executed in $\mathsf{HLT}$. During $\mathsf{KeySwitch}$, one input Ct polynomial expands from $0.5 \cdot \mathcal{B}_{\text{Ct}}^{L}$ to $0.5 \cdot \beta \cdot \mathcal{B}_{\text{Ct}}^{L+k+1}$ for $\mathsf{KeyIP}$ and produces an output Ct of size $\mathcal{B}_{\text{Ct}}^{L}$. Thus, the required on-chip memory for $\mathsf{KeySwitch}$ to eliminate off-chip memory traffic of intermediate Ct is:
\begin{equation}
    \mathcal{M}^{\text{Ct}}_{\mathsf{KeySwitch}}=\mathcal{B}_{\text{Ct}}^{L}+0.5\cdot\beta\cdot\mathcal{B}_{\text{Ct}}^{L+k+1}
\end{equation}

Each $\mathsf{Rot}$ is applied to a Ct consisting of two polynomials. After applying $\mathsf{Automorph}$, one polynomial, $\psi(\mathbf{b})$ is passed to $\mathsf{KeySwitch}$, requiring $\mathcal{M}^{\text{Ct}}_{\mathsf{KeySwitch}}$. In addition, both the original Ct $(\mathbf{a}, \mathbf{b})$ and the $\psi(\mathbf{a})$ must be retained. Therefore, the total on-chip memory required for a single $\mathsf{Rot}$ is
\begin{equation}
    \mathcal{M}^{\text{Ct}}_{\mathsf{Rot}}= \mathcal{M}^{\text{Ct}}_{\mathsf{KeySwitch}}+1.5\cdot\mathcal{B}_{\text{Ct}}^{L}
\end{equation}

For $\mathsf{HLT}$s in Step~1, only one input Ct buffer is required, since $\text{Ct}_A$ and $\text{Ct}_B$ are not reused. For $\mathsf{HLT}$s in Step~2, two input Ct buffers are required, as ($\text{Ct}_{A^{(0)}}$ and $\text{Ct}_{B^{(0)}}$) are reused across multiple iterations. 
In both steps, two output buffers are needed to store the resulting Cts. The multiply-accumulate in $\mathsf{HLT}$ can be computed in place. 
Thus, the on-chip memory requirements for $\mathsf{HLT}$ in each step are:
\begin{equation}
    \mathcal{M}^{\text{Ct}}_{\mathsf{HLT}.s1}=\mathcal{M}^{\text{Ct}}_{\mathsf{Rot}}+3\cdot\mathcal{B}_{\text{Ct}}^{L}
\end{equation}
\begin{equation}
    \mathcal{M}^{\text{Ct}}_{\mathsf{HLT}.s2}=\mathcal{M}^{\text{Ct}}_{\mathsf{Rot}}+4\cdot\mathcal{B}_{\text{Ct}}^{L}
\end{equation}


The memory overhead for the $\mathsf{Mult}$ in Step~2 is minimal compared to $\mathsf{HLT}$. The in-place $\mathsf{Add}$ operation requires an additional buffer to store the accumulated result $\text{Ct}_{AB}$. Therefore, the total on-chip memory requirement for storing Ct during HE MM is given in Eq.~\ref{eq:sram}. Given the matrix dimensions and HE parameter set, all intermediate Cts can be stored on-chip if the available on-chip memory exceeds this requirement.
\begin{equation}
\label{eq:sram}
    \mathcal{M}^{\text{Ct}}_{\text{HE-MM}}=\mathcal{M}^{\text{Ct}}_{\mathsf{HLT}.s2}+\mathcal{B}_{\text{Ct}}^{L}
\end{equation}


\begin{figure*}[t!]
  \includegraphics[width=\textwidth]{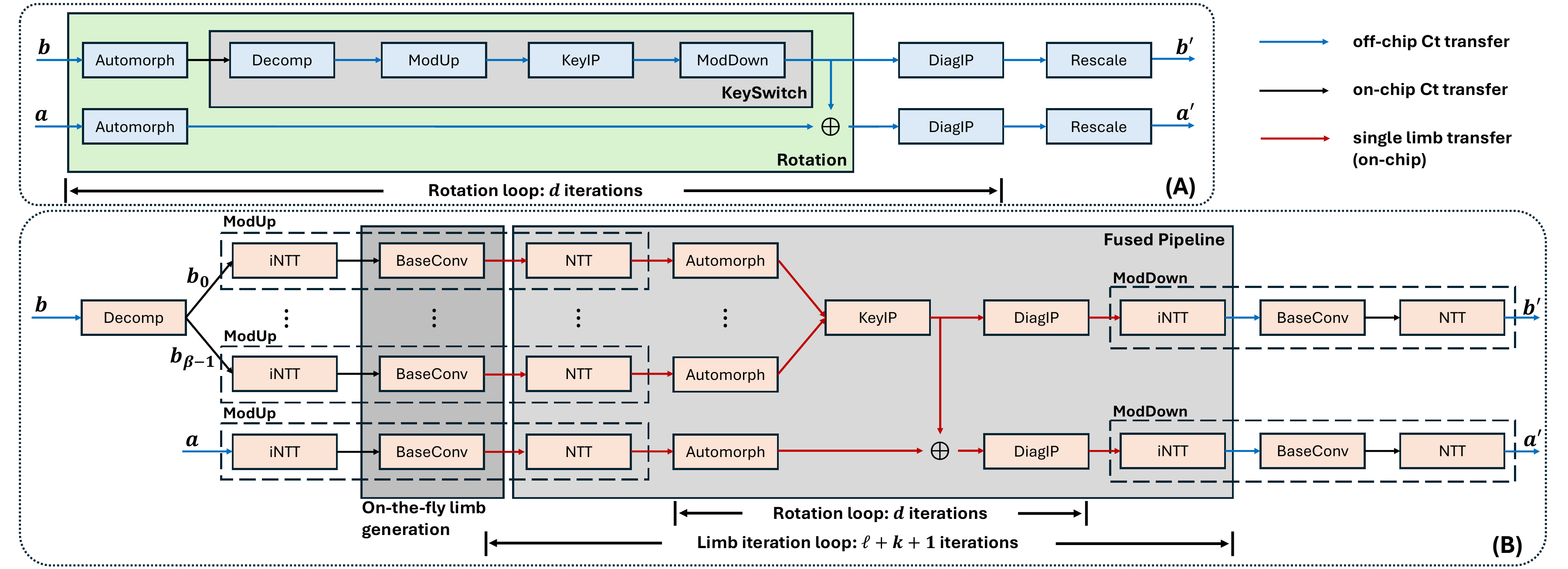}
\caption{For a practical parameter set (i.e., Set-C): (A) Baseline $\mathsf{HLT}$ design with coarse-grained rotation loop and full ciphertext-level datapath, incurring high DRAM traffic.
(B) Proposed memory-optimized $\mathsf{HLT}$ (MO-HLT), an architecture-algorithm co-designed solution, which enables on-the-fly limb generation and sub-operation fusion across $\mathsf{NTT}$, $\mathsf{Automorph}$, $\mathsf{KeyIP}$, $\mathsf{DiagIP}$, and $\mathsf{iNTT}$, drastically reducing SRAM requirement and DRAM access.}
  \label{fig:datapath}
\end{figure*}

\subsubsection{Example Analysis} 

In practice, encrypting a modest-sized matrix (e.g., $64 \times 64$) requires a relatively small HE parameter set with a low polynomial degree, e.g., $N = 2^{13}$. Under such configurations (e.g., Set-A in Table~\ref{tab:param}, Sec.\ref{sec:setup}), each Ct occupies only 0.43\,MB, resulting in a manageable total on-chip memory requirement of approximately 3.6\,MB. This can be comfortably accommodated by SOTA CPUs with a last-level cache (LLC) of 32\,MB per die~\cite{AMDEPYC}.
However, this small parameter set (Set-A) has limited practical utility. Real-world applications generally require larger HE parameter sets (e.g., Set-B and Set-C in Table~\ref{tab:param}) for three main reasons: (1) larger matrices cannot fit within small polynomial degrees; (2) deeper evaluation circuits demand higher Ct levels ($L$) to support consecutive MM operations; and (3) stronger security levels, e.g., 128-bit security, may require larger parameter sizes.

For example, Set-B increases each Ct size to 6.7\,MB, elevating the total on-chip memory requirement to about 61\,MB for HE MM. Even more practical parameter sets, like Set-C, further expand each Ct to 27\,MB, resulting in a total on-chip memory demand of approximately 255\,MB. At this scale, storing two encrypted matrices simultaneously on-chip becomes impractical. Even the requirement for a single $\mathsf{KeySwitch}$ operation ($\mathcal{M}^{\text{Ct}}_{\mathsf{KeySwitch}}$) can exceed available resources. 
Consequently, large parameter sets lead to considerable off-chip memory traffic. Without explicit datapath and memory control, each $\mathsf{KeySwitch}$ can involve hundreds of MBs of off-chip memory access due to repeated reads and writes Cts across sub-operations. Given that each $\mathsf{Rot}$ operation invokes a $\mathsf{KeySwitch}$ and each $\mathsf{HLT}$ typically involves numerous rotations (e.g., 127 for $m=l=n=64$), this can result in tens or even hundreds of GB of off-chip Ct transfers per HLT invocation, creating a severe performance bottleneck.

\subsection{Key Takeaway}
Our analysis highlights that significant off-chip Ct traffic severely limits the scalability of HE MM. The prohibitive memory demands associated with practical HE parameter sets make traditional CPU-based methods infeasible for large-scale HE MM. To overcome this memory bottleneck, Section IV introduces hardware-aware optimizations designed explicitly to minimize off-chip memory accesses and enhance scalability.

\section{Memory-Optimized HLT Datapath}
Motivated by the off-chip traffic bottleneck identified in Sec.~\ref{sec:sram}, we propose a memory-optimized datapath for $\mathsf{HLT}$ (MO-HLT) that reduces on-chip memory requirement to minimize off-chip Ct traffic. Our approach begins with two algorithmic optimizations shown in Algorithm~\ref{algo:decomp_hlt}. The multiply-accumulate step in Line~7 is denoted as $\mathsf{DiagIP}$ (diagonal inner product), encapsulating the $\mathsf{CMult}$ and $\mathsf{Add}$ operations. 
First, sub-operations $\mathsf{Decomp}$, $\mathsf{ModUp}$, and $\mathsf{ModDown}$ are moved outside the rotation loop, which is known as the \textit{hoisting} technique~\cite{de2021does,bossuat2021efficient}. These hoisted sub-operations are shared across multiple $\mathsf{KeySwitch}$ calls and thus executed only once rather than per $\mathsf{Rot}$.
Second, \textit{the $\mathsf{Rescale}$ is merged with $\mathsf{ModDown}$}, enabling a direct modulus reduction from $PQ_\ell$ to $Q_{\ell-1}$, bypassing the intermediate $Q_\ell$~\cite{de2021does,agrawal2023mad}.

\begin{algorithm}[t]
\caption{HLT Represented with Sub-operations}
\label{algo:decomp_hlt}
\textbf{Input:} $\llbracket\bm{m}\rrbracket:=(\bm{a},\bm{b})\in\mathcal{R}_{Q_{\ell}}^2$; Pre-computed Pt diagonals of $U$ and their indices: \{$\bm{u}_{z_t},z_t$\}\par
\textbf{Output:} $\llbracket\bm{m}'\rrbracket:=(\bm{a}',\bm{b}')$ initialized as 0
\begin{algorithmic}[1]
\State{$(\bm{b}_j)_ {j\in[0,\beta)}\leftarrow\mathsf{Decomp}(\bm{b})$}
\State{$(\hat{\bm{a}},\hat{\bm{b}}_j)\leftarrow(\mathsf{ModUp}(\bm{a}),\mathsf{ModUp} (\bm{b}_j))$ for $j\in[0,\beta)$}
\For{$t$ from 0 to $d-1$}
    \State $\hat{\bm{a}}_{\text{rot}}\leftarrow \mathsf{Automorph}(\hat{\bm{a}}, z_t)$
    \State $\hat{\bm{b}}_{\text{rot}}^{(j)}\leftarrow \mathsf{Automorph}(\hat{\bm{b}}_j, z_t)$ for $j\in[0,\beta)$
    \State $(\hat{\bm{u}}, \hat{\bm{v}})\leftarrow\mathsf{KeyIP}(\hat{\bm{b}}_{\text{rot}}^{(j)}, \textbf{evk}_{\text{rot}}^{(z_t)})$
    \State $(\bm{a}',\bm{b}')~+= \bm{u}_{z_t}\cdot(\hat{\bm{u}},\hat{\bm{v}}+\hat{\bm{b}}_{rot})$  
    \Comment{$\mathsf{DiagIP}$}
\EndFor
\State $\llbracket\mathbf{m}'\rrbracket\leftarrow(\mathsf{ModDown}
(\hat{\bm{a}}'),\mathsf{ModDown}(\hat{\bm{b}}'))$
\State \Return $\mathsf{Rescale}(\llbracket\mathbf{m}'\rrbracket)$
\algorithmiccomment{$\mathcal{R}_{Q_{\ell-1}}^2$}
\end{algorithmic}
\end{algorithm}

Integrating algorithmic optimizations above, Fig.~\ref{fig:datapath}(B) illustrates the proposed MO-HLT, an architecture-algorithm co-designed solution featuring two memory-centric optimizations: (1) \textit{fine-grained datapath via sub-operation fusion}, and (2) \textit{on-the-fly limb generation}. These collectively maximize on-chip Ct reuse and minimize off-chip data traffic.

The first optimization, \textit{fine-grained datapath via sub-operation fusion}, improves on-chip Ct reuse by operating at the limb level rather than the full Ct level. As discussed in Sec.~\ref{sec:sram}, full Ct-level reuse is infeasible for large matrices with large HE parameters, due to limited on-chip memory (e.g., 43\,MB of Alveo U280). Instead, we enable reuse across sub-operations by allowing each Ct limb to proceed independently through the datapath, without waiting for all Ct limbs to complete the current stage. This approach significantly reduces both on-chip memory demand and off-chip traffic. Sub-operation fusion is applied within the rotation loop of Algorithm~\ref{algo:decomp_hlt} across $\mathsf{Automorph}$, $\mathsf{KeyIP}$, and $\mathsf{DiagIP}$.

The second optimization, \textit{on-the-fly limb generation}, further reduces on-chip memory pressure by enabling immediate pipelined processing. During $\mathsf{ModUp}$, each newly generated limb from $\mathsf{BaseConv}$ is directly forwarded to $\mathsf{NTT}$ without waiting for the full extended Ct ($\ell + k + 1$ limbs) to be assembled. This allows the $\mathsf{NTT}$ in $\mathsf{ModUp}$ to be seamlessly fused into the pipeline. Similarly, the $\mathsf{iNTT}$ in $\mathsf{ModDown}$ is fused with the preceding $\mathsf{DiagIP}$, reducing off-chip Ct traffic.

Together, these optimizations yield the final fused sub-operation set: {$\mathsf{NTT}, \mathsf{Automorph}, \mathsf{KeyIP}, \mathsf{DiagIP}, \mathsf{iNTT}$}. As shown in Fig.~\ref{fig:datapath}(B), a key architectural change is the reordering of the Ct processing loops. In prior coarse-grained approaches shown in Fig.~\ref{fig:datapath}(A), the outer loop iterates over rotations, and each Ct proceeds through sub-operations collectively at the full-Ct level (i.e., all limbs are processed together for each sub-operation). In contrast, our fine-grained datapath with sub-operation fusion reverses this order: the limb iteration becomes the outer loop, and the rotation loop moves inside. This enables pipelined execution and allows each limb to proceed through the sub-operations independently, improving on-chip Ct reuse.

The proposed MO-HLT reduces on-chip memory requirement by storing only one Ct and $(\beta + 1)$ intermediate limbs to support pipelined processing:
\begin{equation}
  \mathcal{M}^{\text{Ct}}_{\mathsf{MO-HLT}}=\mathcal{B}_{\text{Ct}}^{L}+(\beta+1)\cdot\mathcal{B}_{\text{limb}}
\end{equation}
For Set-C, MO-HLT requires about 29\,MB, easily fitting within typical CPU cache capacities. MO-HLT only reads the input Ct and stores the output Ct, with trivial additional intermediate off-chip Ct transfers during $\mathsf{ModDown}$. The total off-chip traffic for each $\mathsf{HLT}$ is limited to roughly a hundred MB, which is several orders of magnitude smaller than prior CPU-based approaches, making the design highly scalable.
For even larger HE parameter sets (e.g., $N = 2^{17}$, where a Ct can exceed 60\,MB and cannot fit on-chip), MO-HLT remains effective. In such cases, only the unfused sub-operations incur off-chip traffic, while fused sub-operations still benefit from pipelined limb reuse. Overall, MO-HLT significantly reduces both off-chip traffic and on-chip memory demand, effectively addressing the memory bottleneck in practical, large-scale HE MM. Therefore, it is well-suited for on-chip memory-constrained platforms such as CPUs, GPUs, and FPGAs.

\section{HE MM Accelerator - FAME}
\subsection{Overall Architecture}
\begin{figure}[t]
    \centering
    \includegraphics[width=\linewidth]{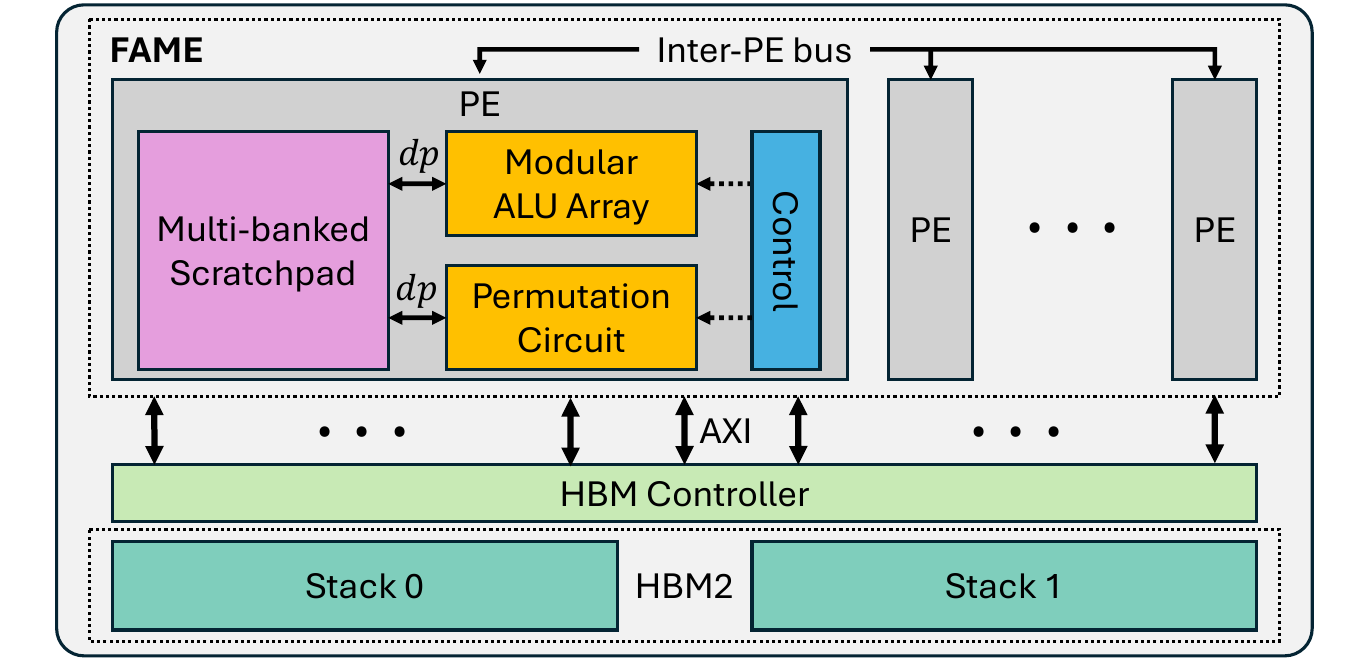}
    \caption{The overall system architecture on FPGA}
    \label{fig:system}
\end{figure}

Fig.~\ref{fig:system} illustrates the system architecture, which consists of the FAME accelerator and two HBM stacks on Alveo U280. To interface with the 32 AXI ports of the HBM, we instantiate 32 asynchronous FIFO pairs for concurrent write and read operations. Each FIFO has a data width of 256 bits, aligning with the bit-width of the AXI interface.
FAME is composed of multiple processing elements (PEs), each designed with a fully pipelined architecture to improve resource utilization and maximize throughput. An inter-PE bus, with a width equal to the data parallelism ($dp$), enables communication between PEs for polynomial transfer.
Prior to computation, data from the read FIFOs is packed into $dp$-wide data lanes and streamed into the multi-banked scratchpad memory within each PE.
FAME features a highly configurable architecture. Key design parameters—including the number of PEs, $dp$ within each PE, and the scratchpad size per PE—can be tuned to accommodate various matrix sizes and HE parameter sets, while adapting to the available FPGA resources.

\subsection{PE Architecture}
Each PE comprises three key components: (1) a modular arithmetic logic unit (ALU) array, (2) a configurable multi-banked scratchpad, and (3) a permutation circuit. 

\subsubsection{Modular ALU Array}
Polynomial computations are reduced to basic modular operations (addition, subtraction, and multiplication). Accordingly, each modular ALU array contains $dp$ modular ALUs, each consisting of one modular adder (configurable for addition or subtraction) and one modular multiplier. The modular multiplier is implemented with a pipelined Barrett reduction design~\cite{kim2019fpga}. We adopt 54-bit RNS prime integers ($\log q$) for efficient utilization of DSPs~\cite{riazi2020heax}.


\begin{figure}[t]
    \centering
    \includegraphics[width=\linewidth]{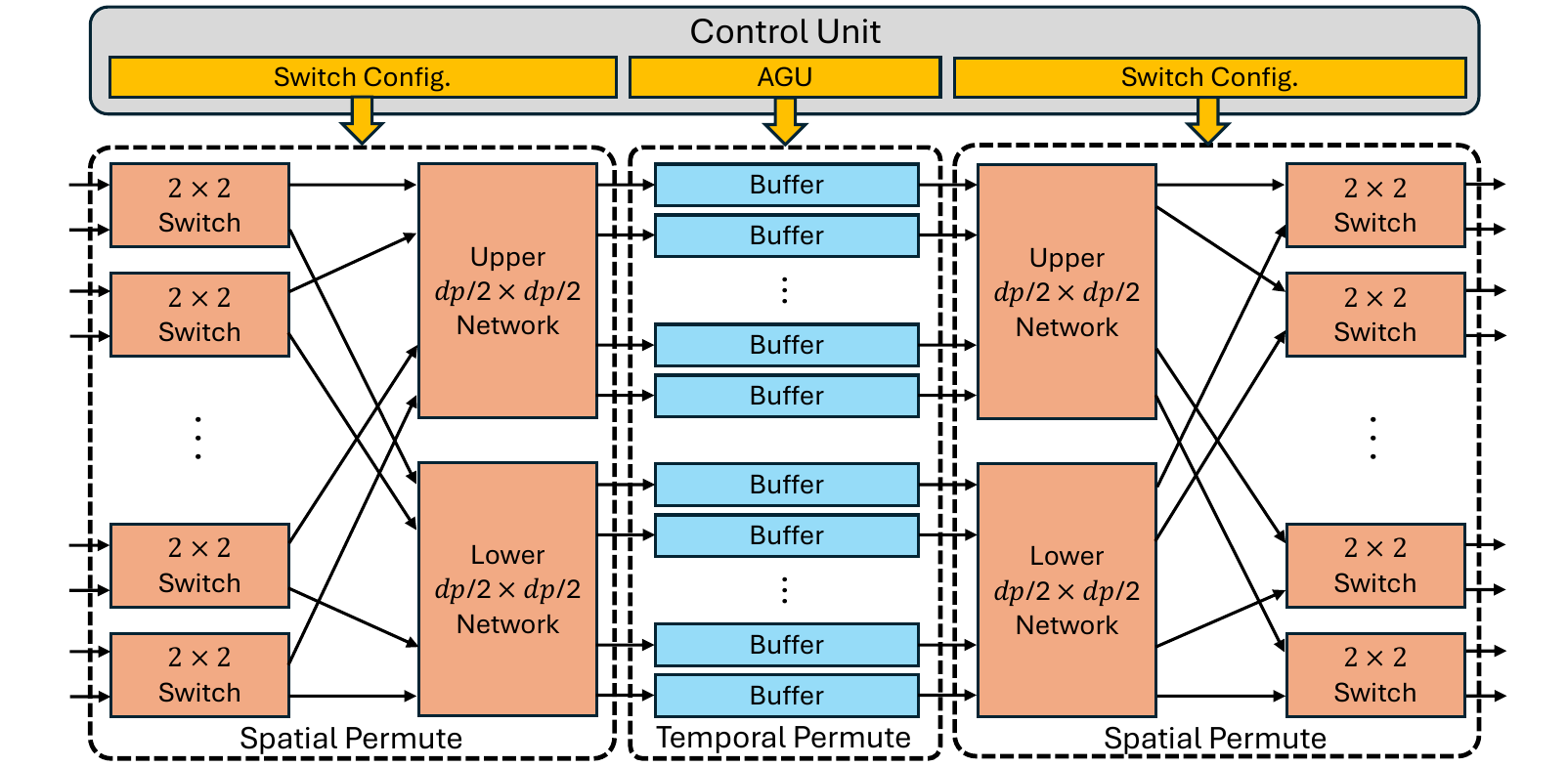}
    \caption{Fully pipelined permutation circuit ($dp$-to-$dp$)}
    \label{fig:spn}
\end{figure}

\begin{figure}[t]
    \centering
    \includegraphics[width=\linewidth]{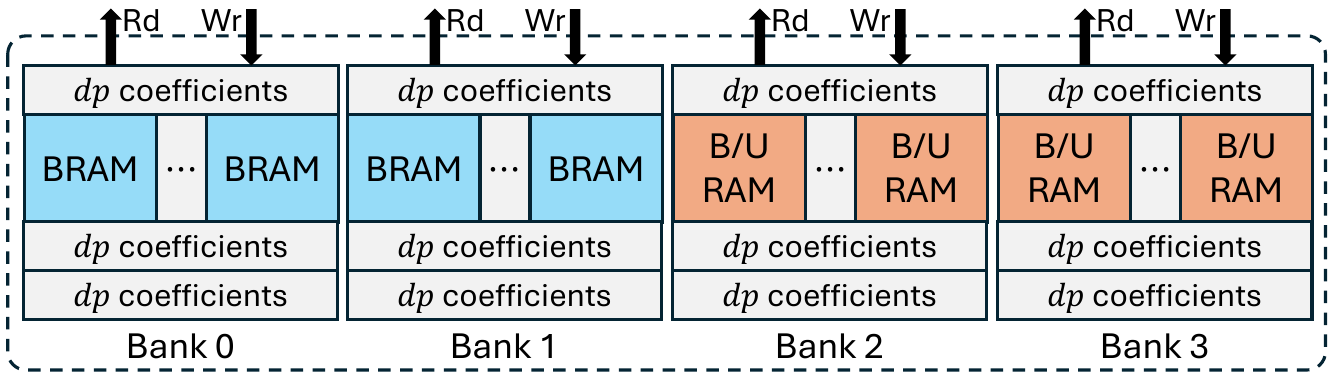}
    \caption{Scratchpad memory organization with BRAM or URAM banks}
    \label{fig:bank}
\end{figure}

\subsubsection{Permutation Circuit}
The permutation circuit, shown in Fig.~\ref{fig:spn}, handles two permutation tasks in HE: (i) radix-2 butterfly permutations for $\mathsf{(i)NTT}$ and (ii) index mapping ($\psi$) of $\mathsf{Automorph}$. 
Our design employs a streaming permutation network (SPN)~\cite{chen2015automatic}, enabling arbitrary parallel coefficient permutations without resorting to a costly crossbar. We redesign its control logic for HE to support $\mathsf{Automorph}$.
The circuit includes three subnetworks: two spatial permutation networks and one temporal permutation network. Each spatial network consists of $\log dp$ stages with $(dp/2)$ 2$\times$2 switches to achieve complete $dp$-to-$dp$ connectivity.
The address generation unit (AGU) coordinates the temporal permutation by controlling $dp$ dual-port buffers. Each buffer stores $N/dp$ coefficients, facilitating coefficient permutations across different processing cycles. The fully pipelined permutation circuit achieves a throughput of $dp$ coefficients per cycle.

\subsubsection{Multi-banked Scratchpad}
The scratchpad memory is organized into four dual-port banks, each with a width of $dp$, as shown in Fig.~\ref{fig:bank}. These banks store various polynomials, including Ct, Pt, $\mathbf{evk}$, and twiddle factors. The scratchpad supports up to four concurrent read and write requests from the modular ALU array, the permutation circuit, and the HBM interface.
To accommodate a large matrix encrypted with a large Ct, two of the banks can be configured and implemented using URAMs.
The depth of each bank is also configurable, allowing the total scratchpad size to be tailored based on the on-chip memory cost model for MO-HLT. This flexibility enables support for a wide range of matrix sizes and HE parameter sets.


\section{Evaluation}
\subsection{Experimental Setup}\label{sec:setup}
We implement FAME using Verilog HDL and map it on Alveo U280. The FPGA has 1,304K LUTs, 2,607K FFs, 43 MB on-chip SRAM, and 9,024 DSPs.
We perform synthesis, place-and-route using Vivado 2023.1. The results are reported after place-and-route. We run RTL simulations to report latency. Latency is defined as the duration from loading the input encrypted matrices from the HBM to storing the output encrypted matrix back into the HBM.

Since prior CPU-based implementations use different schemes, libraries, and platforms, we reimplement several representative approaches~\cite{e2dm,huang2021more,huang2023secure,gao2024secure} using the CKKS scheme with Pyfhel~\cite{ibarrondo2021pyfhel} for fair and consistent benchmarking. 
Experiments are conducted on a high-performance server CPU (Intel Xeon Gold 6326) with 32 cores (64 threads), a 24\,MB cache, and running at 2.9\,GHz. Specifically, the following baseline approaches are implemented for comparison:
\begin{itemize}
    \item E2DM-S~\cite{e2dm}: A method targeting square MM. For general MM $A_{m\times l}\times B_{l\times n}$, matrices are padded to square dimensions $s\times s$, where $s=\max\{m,l,n\}$.
    \item E2DM-R~\cite{e2dm}: An optimized variant of E2DM-S for rectangular MM of the form $A_{m\times l}\times B_{l\times l}$ or $A_{l\times l}\times B_{l\times n}$.
    \item Huang et al.~\cite{huang2023secure}: A pioneering approach supporting arbitrary matrix shapes.
    \item HEGMM-En~\cite{gao2024secure}: A SOTA approach for arbitrary matrix shapes, with improved performance when $m=\min\{m,l,n\}$ or $n=\min\{m,l,n\}$.
\end{itemize}


\subsubsection{HE Parameter Sets and Benchmarks}
We adopt three HE parameter sets (Table~\ref{tab:param}) to support varying matrix sizes. All ensure at least 80-bit security ($\lambda$), with Set-B\&C achieving 128-bit security. Notably, Set-B\&C offer multiple Ct levels, enabling consecutive MMs and supporting real-world applications with deep evaluation circuits.

\begin{table}[t]
    \centering
    \caption{HE Parameter Sets used for Evaluation}
    \begin{tabular}{c|cccccc}
     \toprule
         &  $N$ & $\log Q$ & $L^*$ & $k$ & $\beta$ & $\lambda$\\
    \hline    
       Set-A  & $2^{13}$ & 218 & 4 & 1 & 1 & 80\\
       Set-B  & $2^{15}$ & 855 & 15 & 8 & 2 & 128\\
       Set-C & $2^{16}$ & 1693 & 31 & 12 & 3 & 128\\
    \bottomrule
    \multicolumn{7}{l}{$*$ Fresh Ct levels; supported computation depth}
    \end{tabular}
    \label{tab:param}
\end{table}

We target experiments for general MM with HE. Therefore, we select various matrix shapes for experiments, categorized into four types: Type-I: $n=\min\{m,l,n\}$; Type-II: $l=\min\{m,l,n\}$; Type-III: $m=\min\{m,l,n\}$; Type-IV: square. Our MM benchmarks (denoted as $m$-$l$-$n$) along with their corresponding HE parameter sets are listed in Table~\ref{tab:bench}. 

\begin{table}[t]
    \centering
    \caption{MM benchmarks with diverse sizes and shapes}
    \label{tab:bench}
    \begin{tabular}{l|ccc}
        \toprule
               & Set-A & Set-B & Set-C\\
        \hline
        Type-I & 64-64-16 & 128-128-16 & 160-160-16\\
        Type-II & 64-16-64 & 128-16-128 & 160-16-160\\
        Type-III & 16-64-64 &16-128-128 & 16-160-160\\
        Type-IV & 64-64-64 & 128-128-128 & 160-160-160 \\
        \bottomrule
    \end{tabular}
\vspace{-0.2cm}
\end{table}
\begin{figure*}[t]
    \centering
    \includegraphics[width=\linewidth]{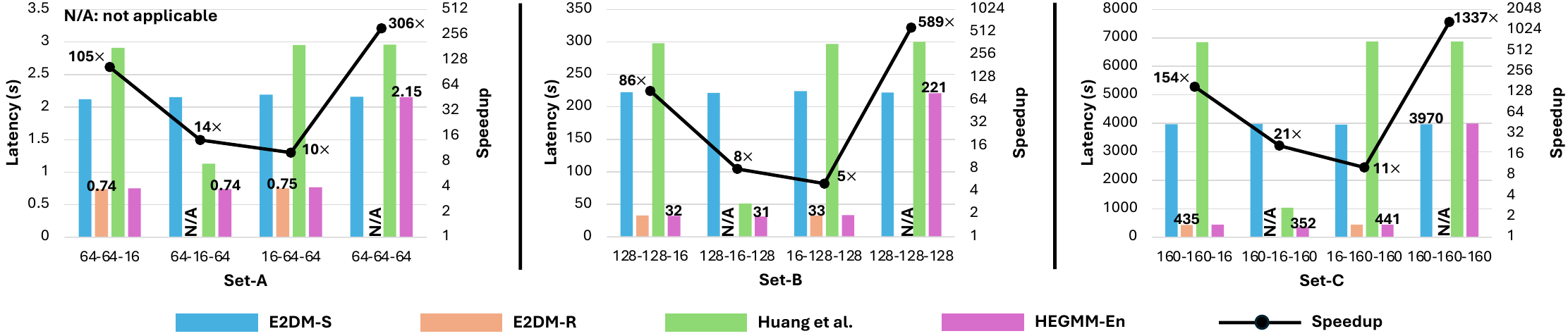}
    \caption{HE MM latency of CPU-based approaches and achieved speedups by FAME over the best CPU result (shown above the corresponding bar)}
    \label{fig:result}
    \vspace{-0.4cm}
\end{figure*}
\subsubsection{FAME Configurations}
We adopt different accelerator configurations tailored to each HE parameter set: FAME-S (Small), FAME-M (Medium), and FAME-L (Large), as detailed in Table~\ref{tab:hardware}. FAME-S is used with Set-A for small matrix sizes, FAME-M with Set-B, and FAME-L with Set-C.

Each configuration features a different scratchpad size to support data reuse for MO-HLT. Under constrained DSP resources, both FAME-S and FAME-M employ 2 PEs, each with a $dp$ of 128. In these configurations, each PE handles the $\mathsf{HLT}$s for one of the input matrices ($A$ or $B$). An inter-PE bus is used to transfer intermediate Ct limbs between PEs after each $\mathsf{HLT}$ in Step~2, enabling the accumulation of the final result.
For Set-C, the on-chip memory is not sufficient to store both $A$ and $B$ simultaneously. Therefore, FAME-L adopts a single-PE configuration with $dp=256$, processing the inputs sequentially while maintaining throughput.

\begin{table}[t]
    \centering
    \caption{FAME configurations and Achieved Frequency}
    \label{tab:hardware}
    \begin{tabular}{c|ccc}
    \toprule
         &  FAME-S & FAME-M & FAME-L \\
         \hline
        \# of PEs & 2 & 2 & 1\\
        \# of Lanes per PE ($dp$) & 128 & 128 & 256 \\
        Scratchpad Size per PE & 864\,KB & 7.6\,MB & 30.4\,MB \\
        Frequency~(MHz) & 350 & 350 & 300\\
    \bottomrule
    \end{tabular}
\end{table}

\subsection{Performance}
Fig.~\ref{fig:result} presents the evaluation results for each CPU-based approach alongside the speedup achieved by FAME. The reported speedup is computed relative to the best-performing CPU baseline for each MM benchmark. The latency of the best CPU implementation is annotated above its corresponding bar.

FAME consistently achieves higher speedups on Type-I and Type-IV benchmarks compared to Type-II and Type-III across all parameter sets. This trend can be attributed to two key factors:
First, FAME exhibits lower latency on Type-I and Type-IV benchmarks because they share the property $m = l$, which results in $d_{U^{\omega^k}} = 2$. Consequently, only two $\mathsf{Rot}$ operations are required for each $\mathsf{HLT}$ involving $U^{\omega^k}$ in Step~2.
Second, CPU-based methods such as E2DM-R~\cite{e2dm} and HEGMM-En~\cite{gao2024secure} implement specialized optimizations for rectangular matrix shapes. In contrast, FAME adopts a unified and general method for handling various matrix shapes, without relying on shape-specific optimizations. As a result, the relative speedup achieved by FAME is lower on Type-II and Type-III benchmarks due to the increased number of rotation operations required.
Among all types, Type-IV consistently yields the highest speedup for FAME across all HE parameter sets, offering a fair comparison against CPU-based approaches under balanced matrix dimensions.

FAME achieves the highest overall speedups in Set-C, compared to the other parameter sets. This aligns with our expectation that the proposed MO-HLT datapath offers greater benefits to reduce off-chip Ct traffic for large benchmarks with a large HE parameter set. 
Notably, it completes the MM of two $160 \times 160$ encrypted matrices in about three seconds—$1337\times$ faster than the best CPU implementation.
On average, FAME delivers a $221\times$ speedup across all benchmarks, implying it can perform roughly 221 consecutive HE MMs in the time a CPU takes to complete just one, assuming sufficient Ct levels are available.
These results highlight FAME’s ability to scale to large-scale consecutive MM efficiently, demonstrating its practicality for real-world, fully encrypted applications.



\subsection{Resource Utilization}
Table~\ref{tab:resource} summarizes the resource usage of each FAME configuration. 
As we scale from FAME-S to FAME-L, memory consumption increases accordingly to support larger matrices and HE parameters. 
In detail, each modular ALU uses 21 DSPs. In addition to the scratchpad, BRAMs are also allocated for temporal buffers in the permutation circuit.

\begin{table}[t]
    \caption{Resource Consumption of FAME configurations}
    \label{tab:resource}
    \centering
    \begin{tabular}{l|ccccc}
        \toprule    
         & DSP & BRAM & URAM & LUT & FF \\
        \toprule
        FAME-S & 5,376 & 1,024 & 0 & 636k & 998k\\
        FAME-M & 5,376 & 640 & 192 & 701k & 1,147k \\
        FAME-L & 5,376 & 3,328 & 672 & 803k & 1,660k \\
        \bottomrule
    \end{tabular}
\end{table}

\subsection{Discussion on Scalability}
This paper considers HE MM with each input matrix encrypted into a single Ct. Thus, the largest matrix size is limited by the Ct polynomial degree.
We demonstrate the scalability of adopting large HE parameters used in practical applications, making it amenable to consecutive MM requiring multiple Ct levels. 
We evaluate FAME on the largest matrix that fits within a single Ct. For even larger matrices, the block MM approach encrypting a matrix with multiple Cts is required.

\section{Related Work}
\textbf{HE Acceleration}
A wide range of HE acceleration efforts have been developed on platforms such as CPUs/GPUs~\cite{helr2,100x,shivdikar2023gme,al2020privft}, FPGAs~\cite{agrawal2023fab,yang2023poseidon,xu2025fast}, and ASICs~\cite{kim2022ark,samardzic2022craterlake,agrawal2023mad}, targeting for various end-to-end HE applications, such as sorting~\cite{hong2021efficient}, image classification~\cite{res20}, and natural language processing~\cite{podschwadt2020classification}. 
However, existing accelerators typically rely on plaintext models, encrypting only the inputs while leaving the model unencrypted. 
To the best of our knowledge, FAME is the first accelerator supporting computations with both inputs encrypted, thus meeting stricter security requirements. 

\textbf{HE MM} is a fundamental kernel for fully encrypted applications, where both input matrices are encrypted. Existing approaches~\cite{mishra2016efficient,e2dm,huang2021more,huang2023secure,gao2024secure} target CPU-based implementations, developing algorithmic optimizations tailored to specific matrix shapes to reduce computational complexity. However, these implementations typically support only small HE parameter sets with limited Ct levels, restricting their applicability to deep evaluation circuits and stronger security requirements.
Due to the lack of scalability in CPU-based solutions, existing methods have struggled to handle HE MM with large HE parameter sets, which limits their applicability to large-scale and consecutive MM workloads. Therefore, we introduce FAME, the first hardware accelerator specifically designed for HE MM. FAME enables efficient and scalable computation under practical configuration and large benchmarks, adopting a unified method applicable across diverse matrix shapes.


\section{Conclusion}
In this work, we presented FAME, the first FPGA-based accelerator for HE MM with both input matrices encrypted. To overcome the performance and scalability limitations of prior CPU-based solutions, FAME leverages the proposed MO-HLT datapath, significantly reducing off-chip traffic through algorithm-architecture co-design.
FAME achieves resource efficiency by minimizing on-chip memory usage. Its configurable architecture supports a wide range of HE parameter sets and matrix dimensions.
Implemented on Alveo U280, FAME demonstrates scalable, low-latency HE MM, paving the way for fully encrypted applications. Future work will explore extending FAME to the block MM approach and real-world workloads requiring consecutive MM. We will also extend FAME with support for shape-specific optimizations.


\section*{Acknowledgment}
This work is supported by the U.S. National Science Foundation (NSF) under grants CSSI-2311870 and SaTC-2104264. Equipment and support from AMD AECG are greatly appreciated. \textbf{Distribution Statement A:} Approved for public release. Distribution is unlimited.

\clearpage
\bibliographystyle{IEEEtran}
\bibliography{references}

\end{document}